# Hydrogen-mediated CVD epitaxy of Graphene on SiC: growth mechanism and atomic configuration

Zouhour Ben Jabra[1], Isabelle Berbezier[1], Adrien Michon[2], Mathieu Koudia[1], Elie Assaf[1], Antoine Ronda[1], Paola Castrucci[3], Maurizio De Crescenzi[3], Holger Vach[4], Mathieu Abel[1]

[1]Aix Marseille University, Université de Toulon, CNRS, IM2NP Marseille, France
[2]CRHEA, Université Côte d'Azur, CNRS, Rue Bernard Gregory, 06560 Valbonne, France
[3]Dipartimento di Fisica, Università di Roma Tor Vergata, via della Ricerca Scientifica 1, 00133 Roma, Italy
[4]LPICM, CNRS, Ecole Polytechnique, IP Paris, Palaiseau, 91128, France

**ABSTRACT**

Despite the large literature focused on the growth of graphene (Gr) on 6H-SiC(0001) by chemical vapour deposition (CVD), some important issues have not been solved and full wafer scale epitaxy of Gr remains challenging, hampering applications in microelectronics. With this study we shed light on the generic mechanism which produces the coexistence of two different types of Gr domains, whose proportion can be carefully controlled by tuning the $H_2$ flow rate. For the first time, we show that the growth of Gr using CVD under $H_2$/Ar flow rate proceeds in two stages. Firstly, the nucleation of free-standing epitaxial graphene on hydrogen (H-Gr) occurs, then H-atoms eventually desorb from either step edges or defects. This gives rise, for $H_2$ flow rate below a critical value, to the formation of (6x6)Gr domains on 6H-SiC(0001). The front of H-desorption progresses proportionally to the reduction of $H_2$. Using a robust and generic X-ray photoelectron spectroscopy (XPS) analysis, we realistically quantify the proportions of H-Gr and (6x6)Gr domains of a Gr film synthetized in any experimental conditions. Scanning tunnelling microscopy supports the XPS measurements. From these results we can deduce that the H- assisted CVD growth of Gr developed here is a unique method to grow fully free-standing H-Gr on the contrary to the method consisting of H-intercalation below epitaxial Gr on buffer layer. These results are of crucial importance for future applications of Gr/SiC(0001) in nanoelectronics, providing the groundwork for the use of Gr as an optimal template layer for Van der Waals homo- and hetero-epitaxy.

Since 2004, the unique properties of graphene (Gr), a single layer of graphite, [1] [2] [3] [4] have attracted a huge number of studies.[5] [6] Instead of graphite presenting a parabolic dispersion at K-point of the Brillouin zone [7], graphene has a planar honeycomb structure and a linear dispersion. [8][9][2] Electrons behave like massless fermions around the Fermi energy level with an unprecedented Fermi velocity. [10][11][12] Graphene was initially obtained by exfoliation from bulk graphite [10] and then transferred onto silicon oxide [4]. This procedure suffers of important electron-phonon scattering at the interface between graphene and the substrate underneath, resulting in a carrier mobility reduction.[13] Alternative substrates were then efficiently explored for the catalytic growth of Gr such as metals (Ni, [14] Pt, [15] Ru, [16] Ir, [17] Pd, [18] Cu [18][19]), III-V alloys, [21] Boron Nitride (h-BN) [22] proving that high quality Gr can be properly transferred on various substrates. However, even if the exfoliation technique is simple, inexpensive and perfectly fitted to fundamental physics, its technological development is not enough when it comes to applications (in particular for microelectronic industry). [23] This because Gr flakes are not large (and homogeneous) enough, despite all the efforts devoted to improve the process for large scale applications [3][24][25].

The growth of graphene by the sublimation process (called graphitization) [26][27][28] was then considered as a technique of choice for potential industrial applications. It is achieved by thermal decomposition of SiC substrates at temperatures around 1600-2000°C under ultra-high vacuum (UHV) [29][30][31] or under Ar atmosphere. [32] The method consists of sublimating silicon atoms from the SiC substrate, leaving behind carbon atoms in a graphene-like layer. About 30% of these carbon atoms are covalently bound to the Si ones underneath at the origin of the (6√3 × 6√3)R30° reconstructed interface layer, called buffer or zero layer, (hereafter referred (6x6)). However, thermal decomposition is not self-limiting and graphene films produced on both Si(0001)- and C(000-1)-terminated surfaces are commonly composed of few sub-micrometer domains with 1-3 MLs thickness. Uniformity on full scale wafers remains challenging. Gr grows much faster on C- face and forms larger domains (~200 nm) of multilayer rotated Gr films (resulting from morphological changes of the SiC surface during Si sublimation) than on Si-face, where domains have <100 nm and a single orientation. [33][34] In order to obtain Gr layers with more uniform thickness, various approaches have been proposed to counteract the Si sublimation rate *e.g.* using simultaneous Si deposition (or using a mixture of Si and $N_2$). The idea is to approach the thermodynamic equilibrium and better control the C-rich (6√3 × 6√3)R30° buffer layer and the resulting graphene thickness. [35][36][37][38][39] A high pressure of Ar was also reported as an efficient way to reduce the Si evaporation since part of the desorbed Si atoms are reflected back to the surface by collision with Ar atoms. [40][41][26] High sublimation temperature was also shown to restructure and uniformize the terraces morphology. Another approach was to decouple the Gr layer from the substrate by H intercalation. [42] Therefore, many attempts were made to decouple the Gr layer from the SiC substrate by H intercalation, so passivating the Si atoms dangling bonds at the SiC interface. [42] This H intercalation process, starting from (6x6)Gr epitaxial layers does not produce full H-Gr layers (as testified by the presence of weak spots representative of the (6√3 × 6√3)R30° buffer layer on LEED patterns).

Gr/metal substrates interface was decoupled by metals intercalation such as: Co, [43] Fe, [44] Au, [45] Ni. [46] Oxygen [47] [48] [49] and hydrogen [50] were also intercalated between Gr and metal substrates. However, Gr decoupled by metals is not suitable for applications in microelectronic and would require the film transfer onto insulating substrates, procedure leading to defect creation.

More recently, Graphene has also been grown using a carbon source [51] either by UHV-molecular beam epitaxy (MBE) [51] or by chemical vapor deposition (CVD) under Ar [52] [53] [54] or $H_2$ atmosphere [55] to meet the requirements of technological applications. SiC that was the first naturally used substrate has many advantages for the direct growth of Gr [22] in particular that of providing reproducible and reliable high-quality Gr layers with a well-controlled and homogeneous thickness at large scale [56] both on Si- and C- faces of SiC(0001). It was reported [53] [57] that depending on the growth temperature and carrier gas pressure, Gr can be grown on top of the buffer layer (henceforth refer to as (6x6)Gr) that has the (6√3 × 6√3)R30° superstructure resulting from covalent bonds with the SiC substrate. The (6x6)Gr obtained by CVD on top of the buffer layer reconstructed (6√3 × 6√3)R30° has a structure very similar to the one of epitaxial Gr obtained by graphitization which has been extensively studied. [53] [52] [54] On the other side, the H-Gr configuration was only reported as the result of the intercalation of $H_2$ below the epitaxial (6x6)Gr using $H_2$ flux at temperatures below 900°C to avoid $H_2$ desorption, [42] [58] [59] while its formation by CVD using C-based gas (i.e. propane) and $H_2$ (or $H_2$/Ar mixture) carrier gas has already been demonstrated by A. Michon et *al.* [56] [55]

In this work, we report the atomic configurations of Gr layers grown by CVD on 6H-SiC(0001) terminated Si, for three different ($H_2$/Ar) carrier gas ratios and we deduce new insights into the Gr growth mechanisms. AFM, STM, LEED and XPS analyses give evidence different structural characteristics of the surfaces depending on the CVD conditions. For the first time, we demonstrate that the complex interplay between growth and etching mechanisms at work in the close to equilibrium conditions used, can produce full coverage of H-Gr or of (6x6)Gr depending on the $H_2$/Ar flux ratio, the transition between the two configurations being controlled only by hydrogen desorption. More precisely, we have found that for a $H_2$/Ar ratio of 9% only (6x6)Gr grows. For $H_2$/Ar increasing ratio, we observe the coexistence H-Gr and (6x6)Gr domains, and finally for 42% ratio a full H-Gr layer is obtained. The amount of H-Gr and the percentage of the coexistence of (6x6)Gr and H-Gr domains have been assessed by x-ray photoelectron spectroscopy (XPS), scanning tunneling microscopy (STM) and low energy electron diffraction (LEED) careful analyses. These investigations indicate that the switch between the two configurations is controlled by hydrogen desorption. The presence of juxtaposed areas of (6x6)Gr and H-Gr, which is underestimated in many studies, has a great influence on the subsequent deposition of Van der Waals heterostructures. [60] [61] [62] [63] [64] In addition, our results show that CVD is a unique technique to grow uniform and homogeneous H-Gr monolayers at full wafer scale and with very low density of defects, that can serve as robust pseudo-substrates for Van der Waals heteroepitaxy.

**RESULTS AND DISCUSSION**

As reported in Methods section, we performed several $H_2$/Ar ratio CVD growths on 6H-SiC(0001) substrate. For all the obtained samples, the atomic force microscopy (AFM) images always present the same features as shown in Figure 1 (the image corresponds to sample A where Gr is obtained at 1550°C, during 15min with 9% of $H_2$). All the samples investigated exhibit large flat terraces with mean widths between 200 nm and 1 µm (figure 1a), separated by multi-layer steps (see line profile in figure 1b). Their height is commonly around 0.75 nm corresponding to tri-layer SiC steps; a few higher ones are also observed with a height of ≈1.5 nm which corresponds to the total SiC crystalline cell (c = 1.51 nm). Some narrower terraces, separated from the larger ones by monoatomic SiC steps (Figure 1c and corresponding line profile in Figure 1d), with 0.25 nm mean height corresponding to single SiC monolayer are also observed. These step heights have already been observed along <1000> direction of the 6H-SiC. [65][51] All these morphological features prove the preservation of the initial topography of the substrate.

Graphene thin film having a carpet-like morphology, completely covers the SiC surface and then cannot be observed at the AFM scale. In the case of a full (6x6)Gr or full H-Gr, there are no changes of the height difference between two neighbour terraces. In the case of coexisting domains, height differences are too small to be visible by AFM (height difference between (6x6)Gr and H-Gr is expected to be 0.154 nm). Few specific irregular step edges and rough terraces morphologies observed on the surface (Figure 1c) are also assigned to the bare SiC substrate underneath. In order to obtain only one monolayer (ML) Gr growth, it is fundamental to suitably dose the percentage of the hydrogen, argon and propane gases. Indeed, the surface morphology strongly depends on the competition between $H_2$ etching of the SiC surface and $C_3H_8$ promoting Gr growth: [66][67] small additions of propane to the hydrogen atmosphere suppresses the etching of SiC, while too large quantities could degrade the surface morphology.[66] Our CVD experimental conditions are very close to the thermodynamic equilibrium ones. This ensures the formation of only one ML of Gr, whatever the growth duration. Gr growth rate mainly depends on $H_2$ flow rate while temperature and total pressure have low effect. Moreover, the introduction of $C_3H_8$ is carefully controlled to avoid any morphological damage.

Atomic structural properties of epitaxial Gr grown on a clean 6H-SiC(0001) are investigated by STM observations and LEED and XPS analyses.

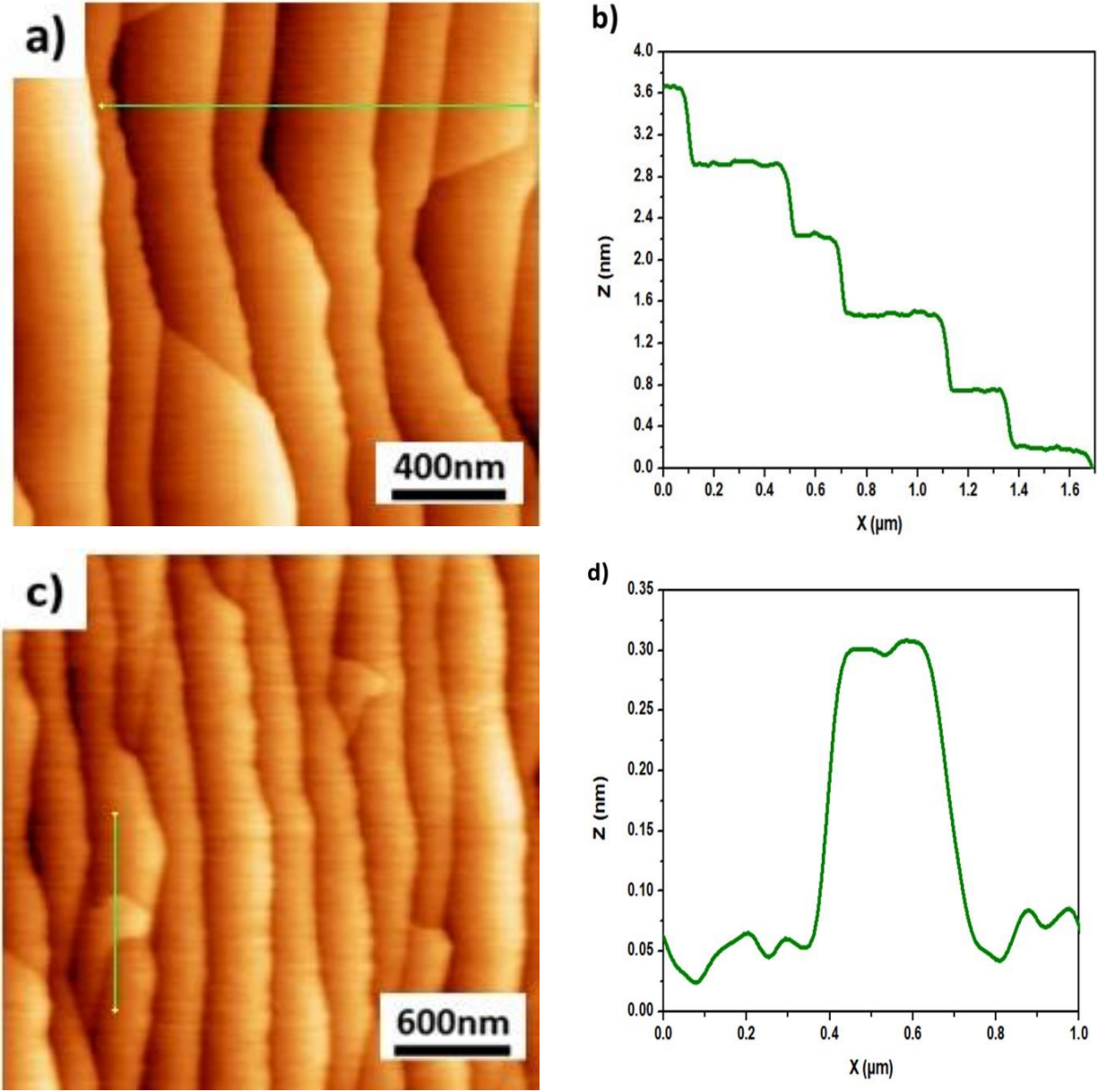

*Figure 1. Graphene epitaxially grown on 6H-SiC (0001) with $H_2$ 9% (sample A). (a) 2 μm × 2 μm AFM image of the Gr/SiC train of steps and terraces; (b) typical line profile of the surface; (c) AFM image of an area with a narrower terrace separated by monolayer SiC steps (0.25 nm height); (d) the corresponding line profile. The z scale for both images is 0- 1.6 nm.*

Systematic investigation of the Gr films was performed on the as grown samples and after annealing. The detailed analysis of three typical samples obtained in different experimental conditions is reported. The first Gr film was obtained in low $H_2$ flux (sample A in Table 1 displayed in Methods). In this condition, incorporation of $H_2$ under the Gr film is very unlikely. Large scale STM images show a train of steps similar to the one observed by AFM separated by steps with heights about 0.75 nm (Figure 2a). Atomic resolution STM images reveal two superimposed honeycomb lattices having periodicities of 3.2 ± 0.1 nm and 0.245 ± 0.2 nm that are ascribed to the (6√3 × 6√3)R30° 6H-SiC supercell [i] (noted (6x6) in the rest of the manuscript for simplicity) and to Gr lattice parameter respectively (Figure 2b). The (6x6) reconstruction has a corrugation height around 0.03 nm which is assigned to the partial interfacial interaction between the C atoms of the first deposited layer (buffer layer) Si atom-terminated 6H-SiC(0001).

The superimposition of the two lattices (6x6)SiC and (1x1)Gr can be appreciated better on a higher magnification STM image (Figure 2c). This feature which has been systematically observed all over the surface of this sample testifies that the buffer layer covers the whole surface. Such STM images have been already reported for epitaxial Gr obtained by sublimation of 6H-SiC(0001) [68] and of 4H-SiC(0001). [58] The main difference is the smoothness and uniformity of the surface which are much better in CVD experimental growth conditions.

The presence of the buffer layer is also confirmed by the LEED pattern (Figure 2d). It reveals spots being the signature of the (1x1)SiC hexagonal lattice (indicated by the pink arrow), of the (1x1)Gr honeycomb network rotated by 30° with respect to SiC one (indicated by the yellow arrow) and higher intensity spots (highlighted by the black arrow) located at vertexes of hexagons and positioned around the Gr spots that are representative of the (6√3 × 6√3)R30° reconstruction. Such patterns have been already reported widely in the literature for Gr grown on buffer layer (that we have named (6x6)Gr). [69] [70] [31] These superstructure spots correspond to the periodicity of the buckling of the buffer layer induced by its strong covalent interaction with the terminating Si-atoms of SiC substrate, as observed by STM. [71] [72] [73]

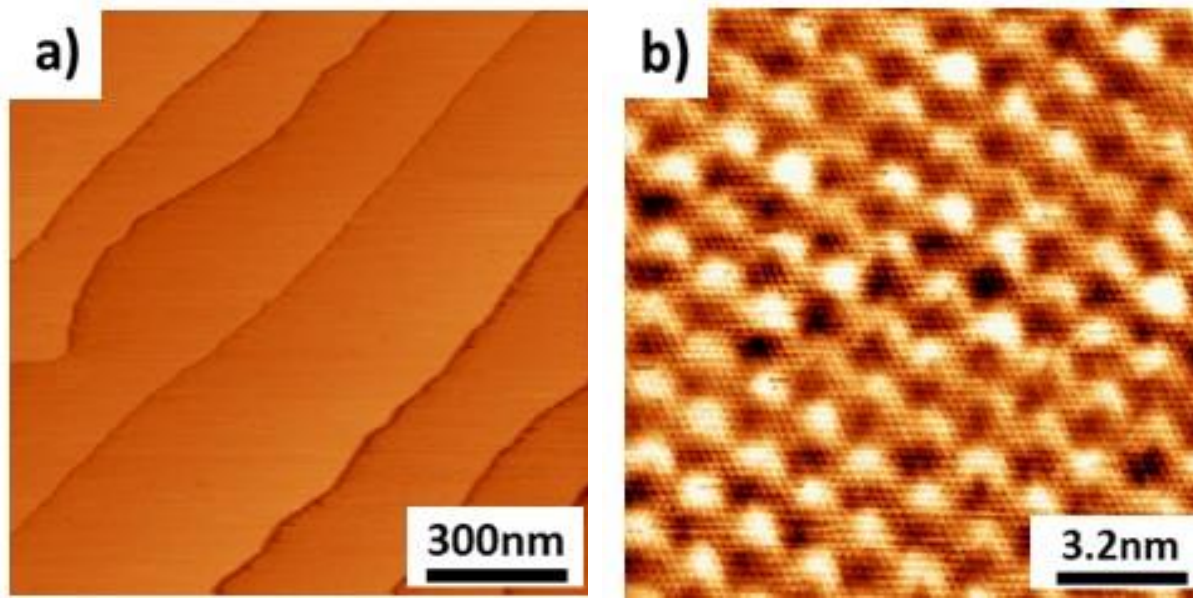

---

[i] (6√3 × 6√3)R30°supercell with 3.2 nm periodicity corresponds to (6x6)Gr pseudo-supercell with periodicity 1.9 ± 0.1 nm (for detailed description see [84] [74]). It also corresponds to (13x13) Gr supercell.

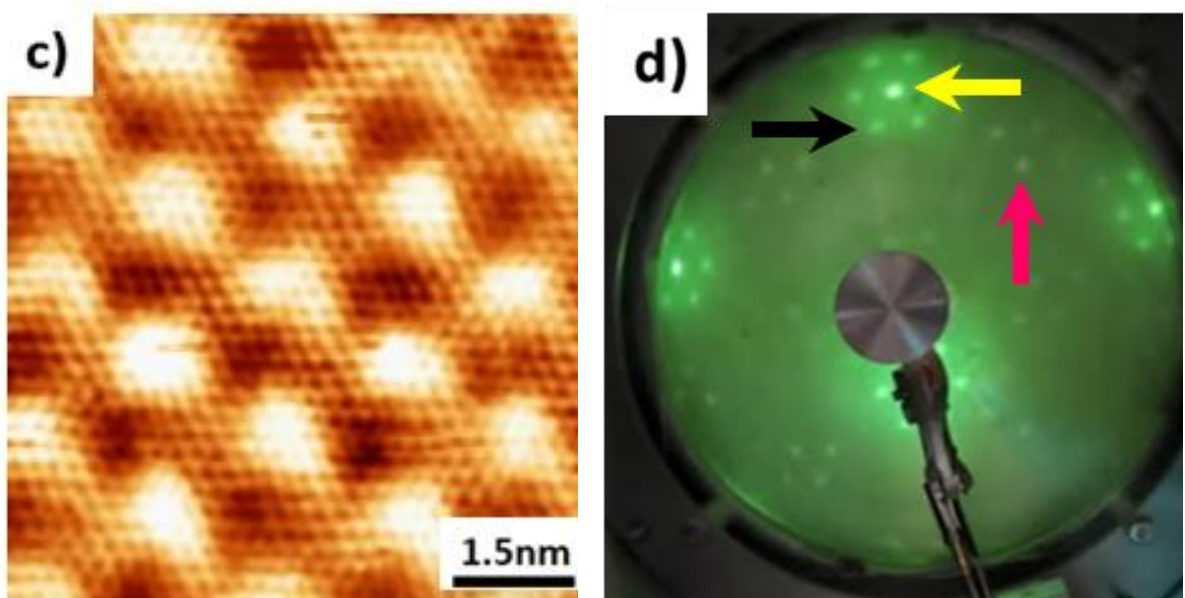

*Figure 2: (a) Large scale STM image showing the train of SiC steps under the Gr surface grown with $H_2$ ratio of 9% (sample A); (b) STM atomic resolution image of the (6x6) and (1x1)Gr superimposed lattices; (c) higher magnification of atomic resolution image of the Gr atomic cell, (d) corresponding LEED pattern : Pink, yellow and black arrows indicate (1x1)SiC, (1x1)Gr and (6√3×6√3)Gr respectively (E= 35 eV). Tunneling parameters: (a) $I_t$= 0.495 nA and $U_{bias}$= -0.412 V; (b) and (c) $I_t$= 0.56 nA and $U_{bias}$= -1.56 V.*

XPS spectra of C1s (Figure 3a) and Si2p (Figure 3b) recorded on the same sample have been fitted and analysed after extraction of the background. The C1s peak was deconvoluted assuming four components already identified in the literature: [74] (a) C-Si bonds of the SiC bulk substrate (corresponding to C in $sp^3$ configuration); (b) the top Gr layer (corresponding to C in the $sp^2$ hybridization); (c) the buffer layer which shares covalent bonds with the substrate; (d) the buffer layer in $sp^2$ configuration (no bonds with the substrate). The two last contributions, named S1 and S2, have been well detailed in [74] and their ratio (S1/S2 integrated area ratio) was estimated around 0.5. [31,74] In these experimental conditions, the best fit of the C1s peak is obtained when the 4 components of the pure (6x6)Gr (see supplementary material for details of the procedure) are located at $E_{SiC}$ =283.4 eV , $E_{Gr}$=284.2 eV, $E_{S1}$=284.6 eV and $E_{S2}$=285.3 eV in very good agreement with literature data. The results show that the sample is fully covered with (6x6)Gr.

Another important issue is the number of deposited (6x6)Gr layers deposited which is a key factor for applications. This quantity is estimated by the quantification of C1s/Si2p area ratio, which is found around 1.55 for this sample. Such ratio excludes the possibility for having more than 1ML and is then refers to a single (6x6)Gr ML. In addition, it confirms the STM observation which evidences a corrugation about 0.03 nm corresponding, according to the literature [70,75] to the first free-standing (6x6)Gr layer. Indeed, by increasing the number of Gr layers, the measured corrugation is expected to lower [71,27,76]. At last high resolution TEM cross-section image of the sample (Figure 3c) evidences the buffer layer and the top (6x6)Gr at the interplanar distances already reported in the literature. Therefore, all these results confirm that in these experimental conditions with low H$_2$ flow rate, a high-quality monolayer of (6x6)Gr is formed in perfect epitaxy with the SiC substrate. The film has few defects and very good homogeneity and uniformity throughout the full-scale sample as testified by large scale TEM cross-section images of the (6x6)Gr layer (see supplementary material). [77]

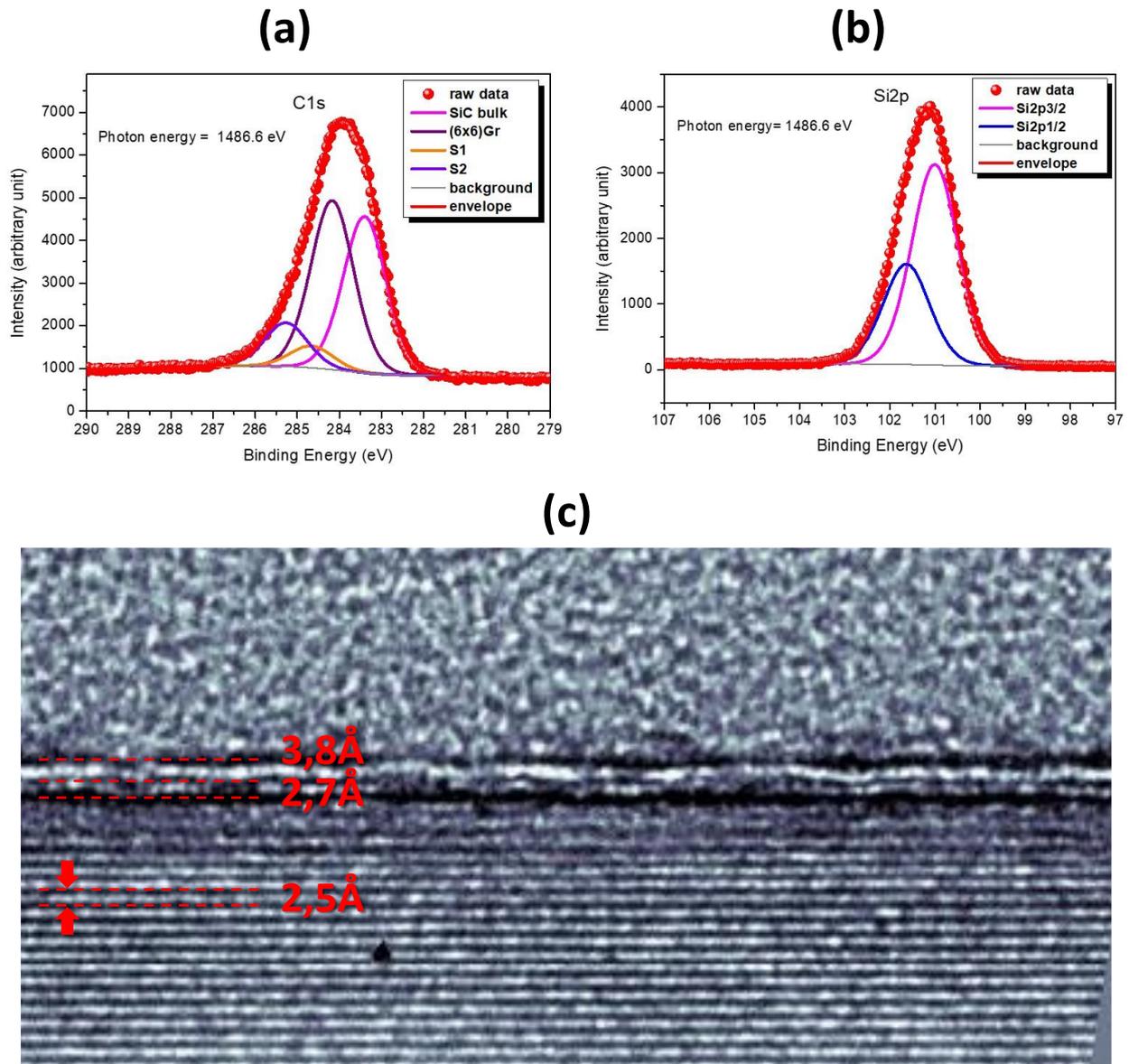

*Figure 3: (a) C1s and (b) Si2p core level XPS spectrum of Gr grown with H$_2$=9% (sample A) and their deconvolution into different carbon components. The different fitted components (solid lines) are labelled in both spectra. The solid red circles refer to the experimental data; (c) High resolution TEM cross-section image of the corresponding sample.*

The second Gr film investigated was obtained with a large H$_2$ ratio ~ 42% (sample C in Table 1). The surface structure observed by STM exhibits regular train of steps with no visible difference on large scale image with the previous sample. However, the atomic resolution images reveal the presence of only one honeycomb structure with 0.25 nm lattice parameter (Figure 4a) which corresponds to the free-standing Gr layer (named H-Gr here). This suggest that in this sample, H is located under the Gr layer interface and saturates the Si dangling bonds as already observed in the literature [42,74,78,58]. Most of the H-Gr layers reported in the literature have been obtained by post-growth hydrogenation of (6x6)Gr, i.e. intercalation of H between the (6x6) buffer layer and the SiC substrate. In these experiments, the buffer layer transforms into a second Gr layer beneath the initial top (6x6)Gr layer. The H diffusion channels pass through step edges and surface crystalline defects. Hydrogenation is commonly incomplete as testified by the (6x6) reconstruction spots always observed on the LEED patterns [42,58]. In our experimental conditions, the complete sample is instead hydrogenated. Indeed, STM measurements do not display any area with (6x6)Gr (Figure 4a) and at higher magnification (Figure 4b), STM images show the uniform intensity of all the atoms of the (1x1)Gr honeycomb network. This is typical of free-standing H-Gr and in stark contrast with graphite where only half of the atoms of the hexagon is visible (triangular cell). Indeed, in graphite, the carbon planes are stacked according to Bernal arrangement, where the second C plane shifted with respect to the first induces that half of the atoms are aligned vertically (resulting in a high intensity), while the others are aligned with respect to the centre of the hexagons of the first plane (low or absent intensity).

In addition, the LEED pattern of this sample exhibits (1×1)Gr and (1×1)SiC intense spots while (6x6) reconstruction is invisible (Figure 4c) confirming the absence of the buffer layer in agreement with STM observations.

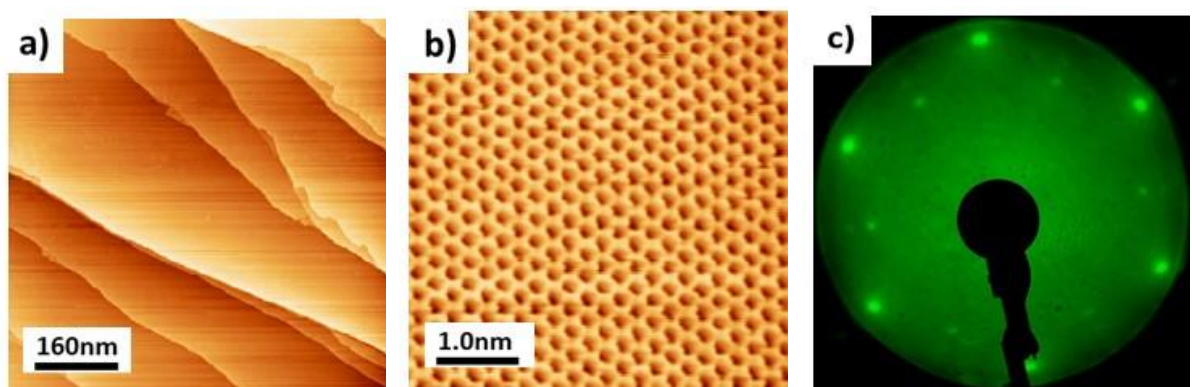

Figure 4: (a) Large scale STM image of Gr grown with $H_2$=42% (sample C) (scanned areas 800 nm × 800 nm) ; (b) higher magnification STM image of (a) which evidences the (1x1) Gr crystalline cell(c) LEED pattern of the same sample at 66 eV. Tunneling parameters are $I_t$=1.36 nA and $U_{bias}$= -0.559 V for both images.

XPS spectra of C1s (Figure 5) recorded on this sample have been analysed in the same conditions as reported above. The C1s peak was deconvoluted assuming two components already identified in the literature for H-Gr [79,80]: (e) the Si-C component referring to the SiC bulk bonded to hydrogen overlayer; (f) the ML Gr with sp² configuration on hydrogen (decoupled from SiC underneath). The best fit is obtained when the C1s experimental peak is deconvoluted into two components with $E_{SiC}$ at 282.5 eV and $E_{Gr}$ at 283.8 eV. A very small tail of the peak observed at high energy has already been observed and is well fitted by replacing Lorentzian profile by a Doniach–Sunjic profile in pseudo-Voigt function.[55] We can note that there is a large shift (0.9 eV) between the SiC signatures of this H-Gr sample and of the previous (6x6)Gr sample, with a lower binding energy when SiC is bound to H than when it is bound to the buffer layer (6x6)Gr. This was explained in the literature by the different band bending occurring at the SiC substrate/Gr interface induced by the different doping types between H-Gr (type P) and (6x6)Gr (type N) [59 74 78]. It was also confirmed by ARPES [32].

As above, we are concerned by the homogeneity and the thickness of the Gr film, that are crucial for many potential applications. The number of deposited layers was evaluated as above, by the quantification of C1s/Si2p area ratio. It is found for this sample around 1.1 which excludes the possibility for having more than 1ML and attests the presence of a single H-Gr ML. This result is consistent with the STM images (reported above) which evidence an equivalent intensity of the six atoms of the (1x1)Gr hexagonal cell as expected for the first free-standing H-Gr layer, otherwise we should have seen a high resolution STM arrangement like to graphite.

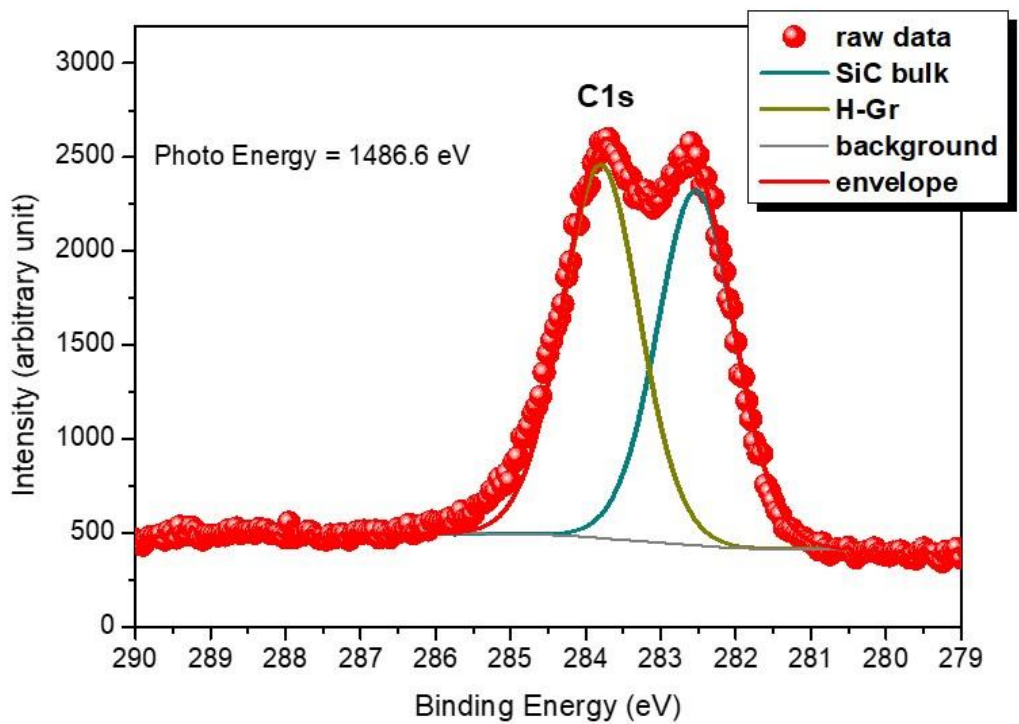

*Figure 5: C1s core level XPS spectrum of graphene grown with $H_2$=42% (sample C). The different fitted components (solid lines) are labelled in the spectrum. The solid red circles refer to the experimental data.*

STM, LEED and XPS show that the use of a high $H_2$ flow rate ensures the production of H-Gr layers of very high quality, good homogeneity on full sample scale and uniform monolayer thickness. They also show that by tuning the $H_2$ flow rate, we can switch from full hydrogenated Gr to (6x6)Gr completely free of hydrogen and lying on the buffer layer, while keeping the layer homogeneous, flat and uniform on the full sample scale.

Let us turn now to intermediate $H_2$/Ar ratios in between the two preceding situations. We have chosen to report the surface structure for $H_2$ ratio ~33% (sample B in Table 1), but similar results were obtained for samples with 10%< $H_2$ ratio <42%, with different proportions of the observed configurations (see quantification and details of the procedure in supplementary material). For sample B, STM observations reveal the juxtaposition of terraces with (6x6)Gr and H-Gr domains on large scales (Figure 6a) whose morphologies are similar to those reported above (sample A for (6×6)Gr terraces and sample C for H-Gr terraces). In other words, there are two time more terraces in sample B with respect to samples A and C and in the latter samples, the size of the terraces corresponds to the mean terrace size of the SiC substrate. The higher magnification images of the interface between these two domains (Figure 6b-c) evidence non-straight step edges with nanofacets following armchair directions for the most part and associated to kinks of 2 nm length, well matching (6x6) unit cells (in size and direction). This observation suggests that the transformation from one configuration to another is made by hopping of (6x6) unit cell following armchair directions. LEED pattern of the sample (Figure 6d) is fully consistent with STM observations and confirms the presence of both configurations with intense graphene spots, surrounded by less intense diffraction spots of the (6x6)Gr. This whole body of evidence proves that the sample is in a metastable (or pseudo-equilibrium) state with a transformation from completely hydrogenated sample to non-hydrogenated sample that can be tuned with the $H_2$ ratio. The distribution of the (6x6)Gr can be estimated from XPS spectra (Figure 7). In this sample, a repartition of 15% H-Gr / 85% (6x6)Gr has been estimated using XPS procedure (supplementary material). This is in good agreement with terraces repartition measured by STM, where a 65 % coverage ratio was assessed for (6x6)Gr on large scale images after identification at high resolution of the two domains.

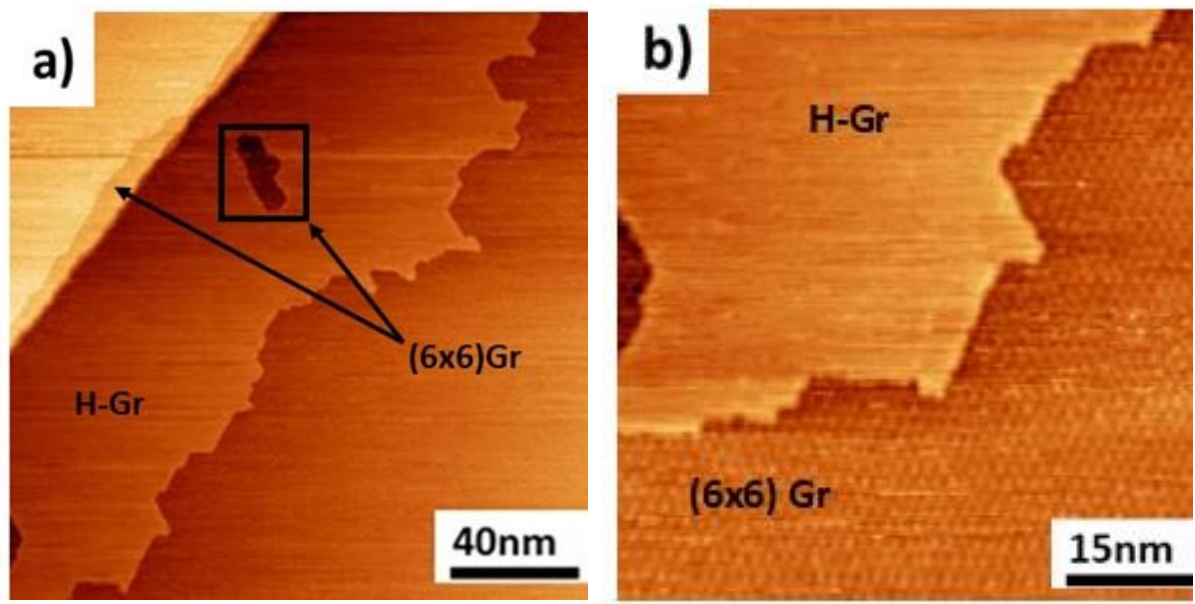

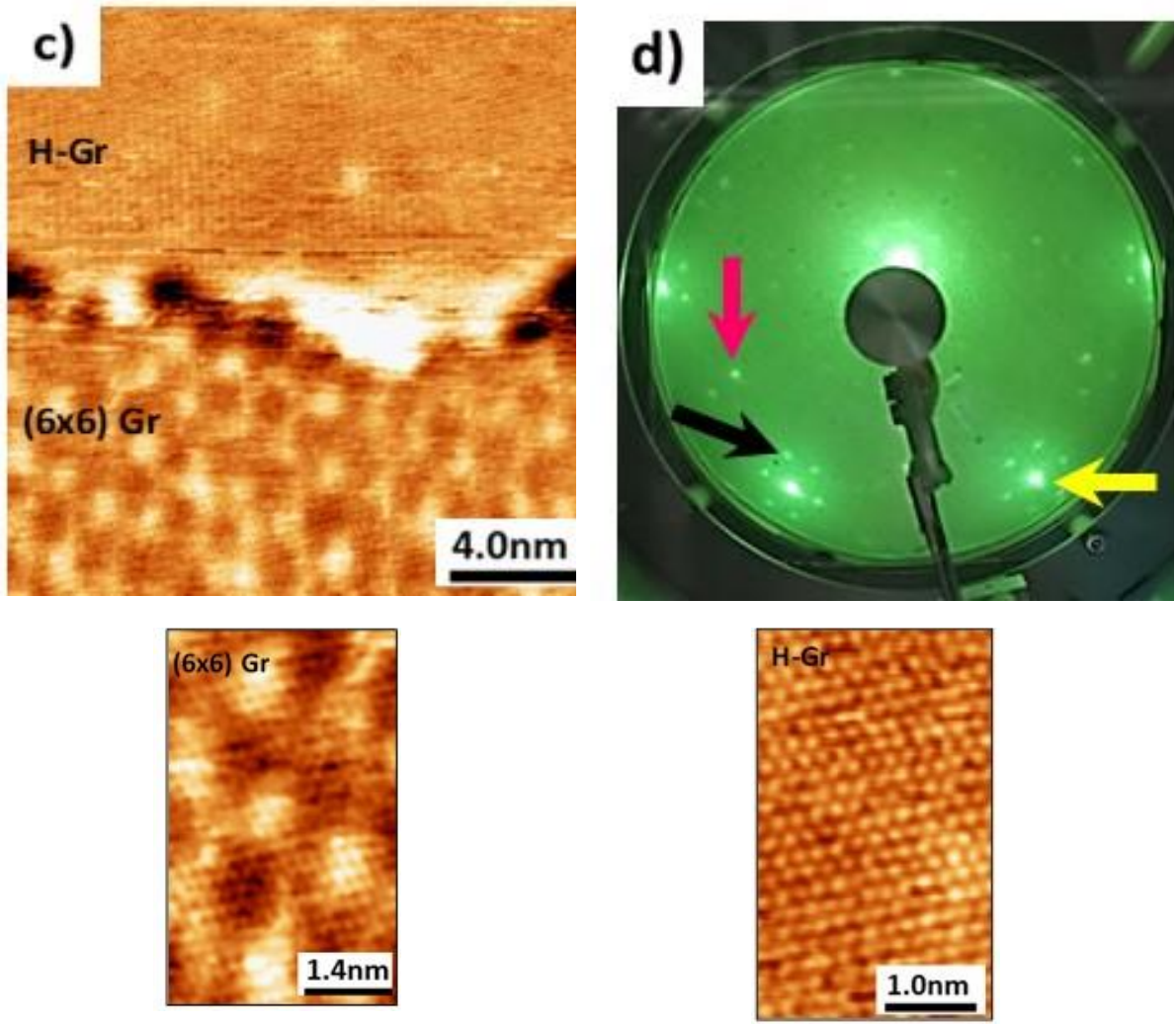

*Figure 6. (a) Large scale STM image of graphene grown with $H_2$ = 33% (sample B); (b) higher magnification image of the interface between H-Gr and (6x6)Gr domains; (c) atomic resolution image of the two configurations and of their interface. The insets represent the atomic resolution of H-Gr (right) and (6x6)Gr (left); (d) LEED pattern of the corresponding sample. Pink, yellow and black arrows indicate (1x1)SiC, (1x1)Gr and (6√3×6√3)Gr respectively. Tunneling parameters for (a) and (b) It=0.317 nA, $U_{bias}$= -0.856 V; (c) and insets: It=0.317 nA, $U_{bias}$= -0.456 V.*

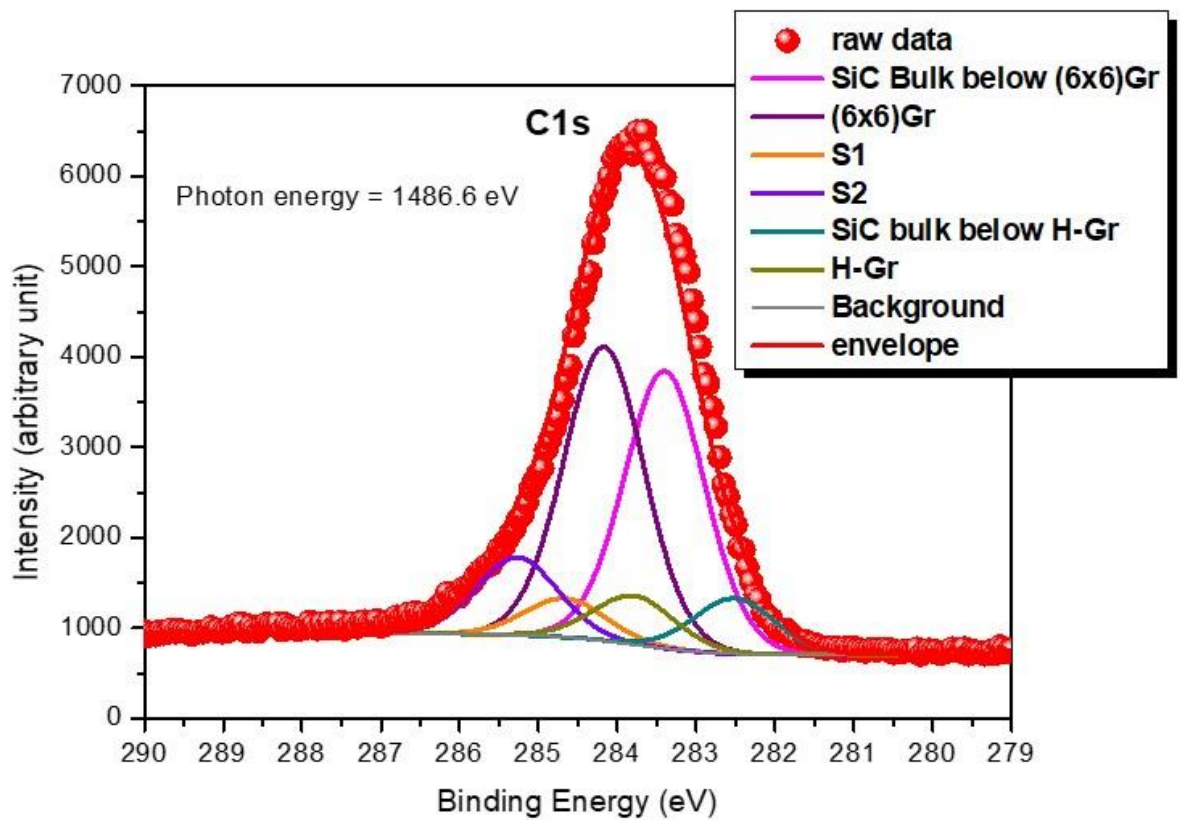

*Figure 7. C1s core level XPS spectrum of Gr grown with $H_2$ =33% (sample B) and their deconvolution into different carbon components. The different fitted components (solid lines) are labelled in the spectrum. The solid red circles refer to the experimental data.*

In summary, AFM observations of Gr/6H-SiC show that whatever the growth conditions are, the surface consists of wide terraces (some hundred nanometers large) separated by a regular train of tri-layer SiC steps (0.75 nm high) with accidentally few monolayer steps (0.25 nm high multiple) or six-layer steps (1.5 nm high). It is obvious that AFM observations cannot visualize the single monolayer Gr which uniformly covers the SiC substrate with a carpet-like morphology. They neither discriminate the surface configurations between H-Gr (hydrogenated graphene) and (6x6)Gr (graphene on buffer layer) domains. In spite of this, we note that AFM is very often used to show the homogeneity of graphene thin films, in particular before deposition of further epitaxial layers, *e.g.* for the growth of Van der Waals heterostructures, whereas Gr atomic structure has a strong influence on both the adsorption energy of many species and their subsequent growth mode. [81][64]

The developed CVD process, which involves a combination of complex processes based on the interaction of $H_2$/Ar and $C_3H_8$ is very robust and reliable. We have shown that it produces an uniform monolayer Gr on full sample scale, whose atomic properties (H-Gr or (6x6)Gr) can be varied as will, by tuning $H_2$/Ar flow rate. This provides exceptional conditions to study the growth mechanism and the transition from one domain to another in detail. In addition, we have shown for the first time, that the proportions of H-Gr and (6x6)Gr can be realistically quantified on large scale by XPS, using a detailed and precise procedure explained in supplementary material. The quantifications obtained have been confirmed by STM assessments.

The growth process can be separated in three chronological stages: (i) sample heating, (ii) Gr growth and (iii) sample cooling. During heating, three mechanisms are at work: Si and C sublimation, SiC etching and passivation by hydrogen. Since a large etching rate favors the surface roughening, Ar is used to reduce the SiC etching (part of the desorbed C and Si atoms collide with argon atoms and return to the surface). Associated to Ar, $H_2$ slowly etches and flattens SiC surface, catalyzing C and Si atoms sublimation accompanied by the formation of Si-H and C-H species. Simultaneously, it also fully covers and passivates the surface. The $H_2$/Ar ratio is then crucial to provide a flat and fully passivated surface during the heating stage. When the growth temperature is reached (>1500°C), the SiC surface is fully hydrogenated. Such high growth temperatures also help to restructure and uniformize the terraces morphology.

When $C_3H_8$ carbon source is introduced in the gas mixture, etching stops and quasi-equilibrium conditions are achieved between etching/sublimation, $H_2$ adsorption and C deposition. In these conditions, the growth is self-limited and only one monolayer of Gr is deposited even for long deposition time. For large proportions of $H_2$, the surface remains fully hydrogenated through all the deposition duration and a single monolayer of Gr uniformly covers the H-terminated SiC substrate. The H-Gr domain is well identified by STM observations that reveal only the Gr atomic cell. LEED and XPS analyses confirm the STM observations.

When reducing the $H_2$ ratio, the proportion of H-Gr decreases and we observe the simultaneous presence of H-Gr and (6X6)Gr domains on juxtaposed terraces. To the best of our knowledge, such morphology has never been observed before. We will consider below the three hypotheses that can explain this morphology: (i) the simultaneous formation of the two domains; (ii) hydrogen intercalation below certain part of the (6x6) domain; (iii) partial desorption of H below the H-Gr domain. Thanks to the systematic observations of the domains and of their boundaries, we can rule out the two first hypotheses and we claim that the (6x6)Gr domains are formed by the partial desorption of H accompanied by the reconstruction of (6x6) unit cells. This mechanism is supported by important aspects of the surface: first, the two juxtaposed domains have large sizes that are not compatible with the concomitant formation of the two domains in homogeneous atmosphere. Second, the step edges between H-Gr and (6x6)Gr domains have mostly armchair orientation, which is the predominant stable configuration and the signature of passivated H-Gr edges [82]. The stability of armchair step edge configuration in presence of hydrogen was explained by both the lower energy of armchair edge (due to triple bonds in the "armrests" while zigzag edges end with strong and expensive dangling bonds) and the weak adsorption of H for armchair (4.36 eV) compared to zigzag (5.36 eV) which stems from the triple bonds in the armchair edge. In addition, the (6x6) kinks generally observed along the steps confirm the transformation from one domain to another, by hopping of (6x6) unit cells, which is well consistent with H-desorption mechanism accompanied by reconstruction of (6x6) unit cells. Third, we have proven that only one H-Gr monolayer is formed, while the intercalation of H below the (6x6)Gr produces two layers of Gr. This confirms that H-Gr is formed first and straight on H-passivated SiC and not on (6x6)Gr. At last, several observations evidence the presence of (6x6)Gr areas fully surrounded by H-Gr domains (see figure 6a for instance). Such a situation can only be explained by the desorption of H from defects on the H-Gr terraces, resulting in the formation of local areas of (6x6)Gr. In presence of C atmosphere, the (6x6) buffer layer is immediately covered by 1ML Gr for thermodynamic stability purpose.

When reducing further $H_2$/Ar ratio, formation of H-Gr is very unlikely. It is either totally inhibited or H-Gr is rapidly converted into (6x6)Gr by H- desorption (due to the lack of $H_2$ molecules) and Gr coverage. The surface exhibits a single (6x6) domain which is well identified by the superimposition of the two structures ((6x6) reconstruction and Gr atomic cell) on STM images. LEED and XPS analyses also confirm the presence of a unique (6x6)Gr domain. It is worth noting that during the cooling stage the adsorption of hydrogen atom from $H_2$ molecules is not favoured due to the cost of $H_2$ dissociation energy, unless the adsorption process should be even more complicated.

**CONCLUSIONS**

By a combined study using a well-controlled growth process and systematic and careful STM/XPS/LEED analyses, we have identified the major steps of the complex Gr growth mechanism in presence of $H_2$, Ar and $C_3H_8$. For the first time, we have shown that the quasi-equilibrium growth process is self-limited and forms a single graphene layer independent of the growth duration. The first stage is the direct growth of a full hydrogenated graphene (H-Gr) over the full wafer scale. Depending on the $H_2$ flux, it eventually transforms into graphene on (6√3 × 6√3)R30° buffer layer ((6x6)Gr) by H-desorption from either step edges or defects on the terraces of 6H-SiC(0001) substrate. Such CVD growth mechanism creates coexisting domains of the two surface structures, which are not distinguishable by AFM observations. It is also a unique process to produce whole layers of hydrogenated graphene or of graphene on buffer layer. It is worth noting that in these conditions, the step edges have an armchair configuration. This could have major impacts, not only from the viewpoint of potential applications (since H-passivated armchair edges have strong influence on the electronic structure of Gr) but also for fundamental science. We also show a robust and general procedure based on XPS data which allows to realistically quantify the proportions of H-Gr and (6x6)Gr domains of a Gr film synthetized in any experimental conditions. Such proportions are of crucial importance for the growth of subsequent layers on top of graphene. We have achieved a precise identification of new Gr configurations, which provides the groundwork for the use of Gr as template layer for Van der Waals heteroepitaxy.

**METHODS**

Graphene samples were grown on Si-terminated face of nominally on-axis Silicon Carbide 6H-SiC(0001) by CVD in a hot wall horizontal growth chamber. The silicon carbide substrates (6H-SiC) are on-axis n-doped (nitrogen) 2'' wafers with typical thickness of ~350μm from Tankeblue. Typical residual offcuts are between 0.05 and 0.2° towards [1–100] direction, as deduced from AFM measurements on annealed substrates under $H_2$ [56]. Before introduction into the growth chamber, the substrates are cut into 1×1cm² pieces and then chemically cleaned with isopropanol. Graphene (Gr) deposition was carried out in a horizontal CVD reactor allowing a homogeneous and reproducible deposition of graphene on the substrates, at a pressure of 800 mbar s using a gas mixture of propane ($C_3H_8$) as the carbon source, and hydrogen ($H_2$) and argon (Ar) as the carrier gases. Growth temperature was varied between 1550 and 1650 °C [56] and durations between 5 and 15 minutes. The $H_2$ flow ratio ($H_2$ flow / total flow) was varied between 9% and 42%. The propane flow ratio was 0.04% or 0.08%. During the temperature ramp to reach the growth temperature and during the cooling down, only the carrier gases ($H_2$+ Ar) were introduced into the growth chamber. The cooling down to room temperature begin at a rate about 4°C/min.

In this study we report the results obtained on three representative samples obtained with different $H_2$ flow ratios: A) 9%, B) 33% and C) 42%. The other experimental conditions listed in Table 1, are expected to have almost no influence on the Gr/substrate interactions as already reported.

| Sample | t (min) | % $H_2$ | T (°C) | $C_3H_8$ (sccm) | Gr/substrate interactions |
|--------|---------|---------|--------|-----------------|---------------------------|
| A | 15 | 9 | 1550 | 0.08% | Gr on buffer layer |
| B | 5 | 33 | 1600 | 0.04% | Partly hydrogenated |
| C | 5 | 42 | 1550 | 0.04% | Complete hydrogenation |

***TABLE 1:** Growth parameters used for the samples studied in this work. All the samples were prepared under the same pressure (800 mbars). The last column gives the Gr/substrate interface configuration.*

Morphological characterization of the samples was investigated by near field microscopy: atomic force microscopy (AFM) in tapping mode in air using Silicon tip and scanning tunnelling microscopy (STM) using electrochemically etched tungsten tips in UHV conditions. The UHV chamber was equipped with low energy electron diffraction (LEED), monochromatic X-ray photoemission spectroscopy (XPS) and STM (Omicron). STM observations were carried out at room temperature using constant current mode with $I_t$ = 0.3 nA and $V_t$ - between 0.4 and 1.5 V. A photon energy of 1486.6 eV (Al-Ka) for XPS measurements was used. The photoelectrons are analysed with an Omicron EA 125 energy analyser. AFM and STM data were analysed by WSxM [83] and ImageJ software. The XPS spectra were fitted using CasaXPS software. Cross-section transmission electron microscopy (TEM) samples were prepared using a FEI Helios 600 Dual Beam $Ga^+$ focus ion beam (FIB). High resolution TEM (HR-TEM) observations were performed using a FEI Titan 80-300 Cs corrected microscope, operating at 200 keV.